\documentclass[preprint,preprintnumbers,amsmath,amssymb]{revtex4}
\usepackage{amsfonts}
\usepackage{amssymb}
\usepackage{latexsym}
\usepackage{graphicx}
\usepackage{psfrag}
\usepackage{lscape}
\newcommand{\al}{\alpha}

\newcommand{\Si}{\Sigma}

\newcommand{\De}{\Delta}

\newcommand{\rar}{\rightarrow}

\newcommand{\non}{\nonumber}
\begin{document}

\title{Critical charges of simple Coulomb molecular systems: One-two electron case}

\author{A.V.~Turbiner}
\email{turbiner@nucleares.unam.mx}
\author{H.~Medel Cobaxin}
\email{medel@nucleares.unam.mx}
\affiliation{Instituto de Ciencias Nucleares, Universidad Nacional
Aut\'onoma de M\'exico, Apartado Postal 70-543, 04510 M\'exico,
D.F., Mexico}


\begin{abstract}
Let us consider some Coulomb systems of several infinitely massive centers of
charge Z and one-two electrons: $(Z,e)$, $(2Z,e)$, $(3Z,e)$, $(4Z,e)$,
$(2Z,e,e)$, $(3Z,e,e)$. It is shown that the physical, integer charges
$Z=1,2,\ldots$ do not play a distinguished role in total energy and equilibrium
distance of a system giving no indication to a charge quantization.

By definition, a critical charge $Z_{cr}$ for a given Coulomb
system $(nZ,e)$ or $(nZ,e,e)$ is a charge which separates a domain of the
existence of bound states from a domain of unbound ones (continuum).
For all above-mentioned systems critical charges $Z_c$ as well as equilibrium geometrical configurations are found.
For all studied systems there was obtained an indication to
a square-root singularity at $Z=Z_{cr}$.
\end{abstract}


\maketitle

\section{\protect\bigskip introduction}

One of the basic observations of fundamental physics is the quantization
of electric charges of elementary particles and nuclei.
The electric charges of electron and proton have opposite signs and their
values coincide, neutron has zero electric charge, any nuclear electric
charge is equal to proton charge multiplied by integer number. This
observation is supported experimentally, gets its justification in elementary
particle theory and nuclear physics. A natural question to ask is there any
indication to such a quantization in atomic-molecular physics. In classical electrostatics the stable configurations of point charges are absent (the
Earnshaw's theorem), zero charge looks like as a singular point where the nature of interaction changes from repulsion to attraction. Usually, at a singular charge the whole or a part of the potential vanishes. In non-relativistic quantum electrodynamics these singular charges undoubtedly continue to exist, however, a new phenomenon occurs - there are some critical charges which separate the domain of the existence of the bound states from the domain of non-existence, although the nature of potential remains unchanged. In some cases a system gets bound at a critical charge with polynomially-decaying eigenfunctions at large distances unlike standard exponentially-decaying eigenfunctions. To the best of our knowledge this phenomenon was observed for the first time for the Helium-like system $(Zee)$. It was named
as "the level hits (kicks) continuum", or as "the zero-energy state", or as "the level on the threshold of continuum". Probably, two simplest examples where such a phenomenon occurs are the P\"oschl-Teller potential and the Yukawa potential.

In this paper we consider a Coulomb system of a number of infinitely massive centers of the same charge Z and one-two electrons assuming that the charge $Z$ is a real parameter. The main goal of this paper is to explore a question: are integer (physical) charges $Z$ special in some sense when the total energy is studied? Another goal is to find the domain(s) in $Z$ where the system has at least one bound state. We intend to find the critical charges $Z_{cr}$ which separate the domains of existence/non-existence of bound state. The study is made in framework of the non-relativistic quantum mechanics.

\section{Generalities}

Let us consider the Coulomb molecular system which consists of $n$ fixed
charges $Z$ and $k$ electrons, $(nZ, ke)$. The Hamiltonian which describes
this system is written as follows
\begin{equation}
\label{Hnk}
{\cal H}\ =\ -\frac{1}{2} \sum_{a=1}^k \Delta_a \ +\ \sum_{i<j} \frac{Z^2}{R_{ij}}
\ -\ \sum_{i=1}^{n}\sum_{a=1}^{k} \frac{Z}{r_{ia}}  \ +\ \sum_{a<b}^k \frac{1}{r_{ab}}
\ ,
\end{equation}
in a.u., where $R_{ij}$ is the distance between charge centers $i$ and $j$,
$r_{ia}$ is the distance from $a$th electron to $i$th charge center, $r_{ab}$ is the distance between electrons $a$ and $b$, and $k=1,2$. If $Z=1$, the Hamiltonian describes the system of $n$ protons and one-two electrons in the Born-Oppenheimer approximation of the zero order (the protons are considered to be infinitely-massive).  There are three important particular cases.

(1) Atomic-type case, $n=1$. The Hamiltonian (\ref{Hnk}) gets a form
\begin{equation}
\label{H1k}
 {\cal H}\ =\ -\frac{1}{2} \sum_{a=1}^k \Delta_a\ -\
 Z\sum_{a=1}^{k} \frac{1}{r_{a}}  \ +\ \sum_{a<b}^k \frac{1}{r_{ab}}\ ,
\end{equation}
where $r_{a}$ is the distance between $i$th electron and the
center. At $k=1$ (one-electron case) we get a hydrogen-like ion, its
spectrum is known
\begin{equation}
\label{H-atom}
    E_{N}(Z)\ =\ -\frac{Z^2}{2 N^2}\ ,
\end{equation}
where $N$ is the principal quantum number, $N=1,2,\ldots$. Discrete spectra is
infinite for any $Z>0$. Critical point is at $Z_{cr}=0$. Nature of
this critical point is of quite obvious - it is a singular point of
the differential equation, at $Z=0$ the potential vanishes. It is
worth noting that $E_{N}(Z)$ has no singularities at finite $Z$ having
the pole of the second order at $Z=\infty$.  For $n>1$ it is evident
from physical point of view that for small $Z$ the system is unbound
but gets bound for sufficiently large $Z$. Hence, there exists some
$Z=Z_{cr}$. It seems established that $Z_{cr}(n=2) \approx 0.91$
(see e.g. \cite{Baker:1990, Zamastil:2010} and references therein) and
$Z_{cr}(n=3) \lesssim 2.1$ (see e.g. \cite{kais}). Making a rescaling
of the Hamiltonian (\ref{H1k}), $r \rar \frac{r}{Z}$ we get
the Hamiltonian in the form
\begin{equation}
\label{H1k-r}
 \tilde {\cal H}\ =\ -\frac{1}{2} \sum_{a=1}^k \Delta_a\ -\
 \sum_{a=1}^{k} \frac{1}{r_{a}}  \ +\
 \frac{1}{Z}\sum_{a<b}^k \frac{1}{r_{ab}}\ ,
\end{equation}
and arrive immediately at the conclusion that the energy of the
bound state has the second order pole at $Z=\infty$. In general,
in the domain $[Z_{cr}, \infty)$ the ground state energy $E(Z)$ is smooth
monotonous function of $Z$ without any indication to a charge quantization.

(2) One-electron, molecular-type case, $k=1$. The Hamiltonian has a form
(\ref{Hnk}) without the last sum. It is evident that for small $Z$ the system
is bound and one critical charge coincides with the singular point of
the Hamiltonian $Z_{cr}=0$, where the potential vanishes. For large $Z$ the
Coulomb repulsion of charged centers gets larger than the attraction
of the electron to them and a system definitely becomes unbound. Hence,
the second critical charge at finite $Z_{cr}>0$ must exist. Our goal is
to find this critical charge for $n=2,3,4$.

(3) Two-electron, molecular-type case, $k=2$. The Hamiltonian
(\ref{Hnk}) gets a form
\begin{equation}
  \label{Hn2}
  {\cal H}\ =\ -\frac{1}{2} (\De_1+\De_2) \ +\ Z^2\sum_{i<j}^n \frac{1}{R_{ij}}
  \ -\ Z\sum_{i=1}^{n} \left(\frac{1}{r_{i1}}+\frac{1}{r_{i2}}\right) \ +\
  \frac{1}{r_{12}}\ .
\end{equation}
From physical point of view it seems evident that a system is not
bound for large $Z$ as well as for $Z \leq 0$. Thus, there must exist
two critical points: one has be near zero, $Z \sim 0$ and another one
has to be at finite $Z$. None of them is of a type of singularity of
the operator (\ref{Hn2}). Our goal is to find this critical charge for
$n=2,3$.

It is necessary to introduce a formal definition of the critical charge
$Z=Z_{cr}$ for molecular system. It is natural to do it in the Born-Oppenheimer
approximation of the zero order when $Z$ charges are assumed to be fixed.
In the case of the existence of a bound state the potential curve
$E_{total}=E_t(R)$ has a minimum at finite internuclear distance $R = R_{eq}$.
If the bound state is stable the potential energy at infinite intercenter
distance is larger at $R_{eq}$. Otherwise, the bound state can be metastable globally: the system can decay and the potential energy at infinite intercenter distance is smaller than at minimum. It implies the existence of the
maximum on the potential curve at some finite $R > R_{eq}$. In the case of non-existence of a bound state the potential curve has no minimum at finite distances. Hence, at critical charge $Z=Z_{cr}$ the potential curve has a saddle point at a finite distance $R$.

\newpage

\section{One-electron molecular systems}

\subsection{Two center case $(2Z,e)$}

\begin{figure}[h]
\includegraphics[angle=00, width=0.25\textwidth]{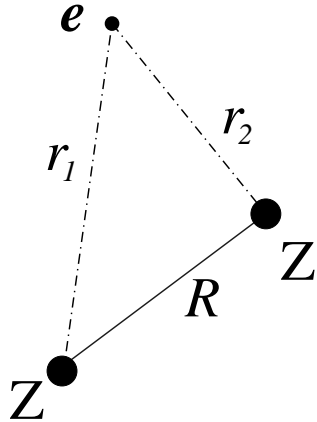}
\caption{Geometrical setting for $(2Z,e)$ system}
\end{figure}

\vskip 1cm

It is well known that at $Z=1$ there exists the molecular hydrogenic ion
H$_2^+$, while at $Z=2$ the molecular helium ion {\rm He}$_2^{3+}$ does
not exist. Hence, the critical charge $Z_{cr}$ has to be in the range
$1 < Z < 2$.


In order to calculate the total energy of $(2Z,e)$-system vs $R$ as a function of charge $Z$ we use the variational method. As a trial function is taken linear superposition of the Heitler-London, Hund-Mulliken and Guillemin-Zener functions
(see \cite{PR}).

{\bf I.} The Heitler-London function.
\[
 \psi_1\ =\ e^{ - \al_1 Z(r_1+r_2)}
\]
where $\al_1$ is variational parameter. It is worth mentioning the
potential, for which the function $\Psi_1$ is exact ground state wavefunction,
\[
V_{trial}^{(1)} \ =\  -2 \al_1 Z\left( \frac{1}{r_1}+ \frac{1}{r_2}\right)
 + 2{\al_1}^2 Z^2 {\vec{n_1}}\cdot{\vec{n_2}}\ ,\ E_1 = 0\ ,
\]
reproduces both Coulomb singularities and at $\al_1=1$ even their
residues. The parameter $\al_1 \ne 1$ makes sense (anti)screening of
the nuclear charges. It is well-known for $Z=1$ that the Heitler-London function
describes small internuclear distances and can give a significant
contribution near equilibrium, at $R \approx R_{eq}$. It mimics a
coherent interaction of the electron with charged centers. It seems evident it holds
for $Z \neq 1$.

{\bf II.} The Hund-Mulliken function.

\[
\psi_2 = \left(e^{-\al_2 Z r_1} + e^{-\al_2 Z r_2}\right)
\]
where $\al_2$ is variational parameter. It describes incoherent
interaction of the electron with charged centers. This function gives
a significant contribution for large internuclear distances.

In order to interpolate between domains $R \simeq R_{eq}$ and $R\gg  R_{eq}$,
we use two interpolating functions.

{\bf III-1.} The Guillemin-Zener function

It is the simplest non-linear interpolation between $\psi_1$ and $\psi_2$ or, saying differently, between small and large internuclear distances,
\[
\psi_{3_1} = \left(e^{-\al_3 Z r_1 -\al_4 Z r_2} +
 e^{-\al_3 Z r_2 -\al_4 Z r_1}\right)
\]
where $\al_3, \al_4$ are variational parameters. If
\begin{itemize}
\item $\al_3 = \al_4 \quad \mbox{ then} \quad \psi_{3_1}
\to \psi_1$
\item $\al_4 = 0 \quad \mbox{ then} \quad
\psi_{3_1} \to \psi_2$
\end{itemize}

{\bf III-2.} Linear Interpolation

\[
\psi_{3_2} = A_1 \psi_1 + A_2 \psi_2
\]

{\bf IV.} Superposition of the two kinds of interpolation
\begin{equation}
\label{trial-2Ze}
    \psi_4 = A_{3_1} \psi_{3_1} + A_{3_2} \psi_{3_2}\ .
\end{equation}

\begin{figure}[ht]
\centering
\psfrag{Total Energy as function of the Charge}{}
\psfrag{Charge}{\huge{$Z$}}
\psfrag{Energy}{\huge{Total Energy}}
\psfrag{Zpe}{(Z,e)}
\psfrag{ZpZpe}{(Z,Z,e)}
\psfrag{ZpZpZpe}{(Z,Z,Z,e)}
\psfrag{ZpZpZpZpe}{(Z,Z,Z,Z,e)}
\includegraphics[angle=-90, width=1.0\textwidth]{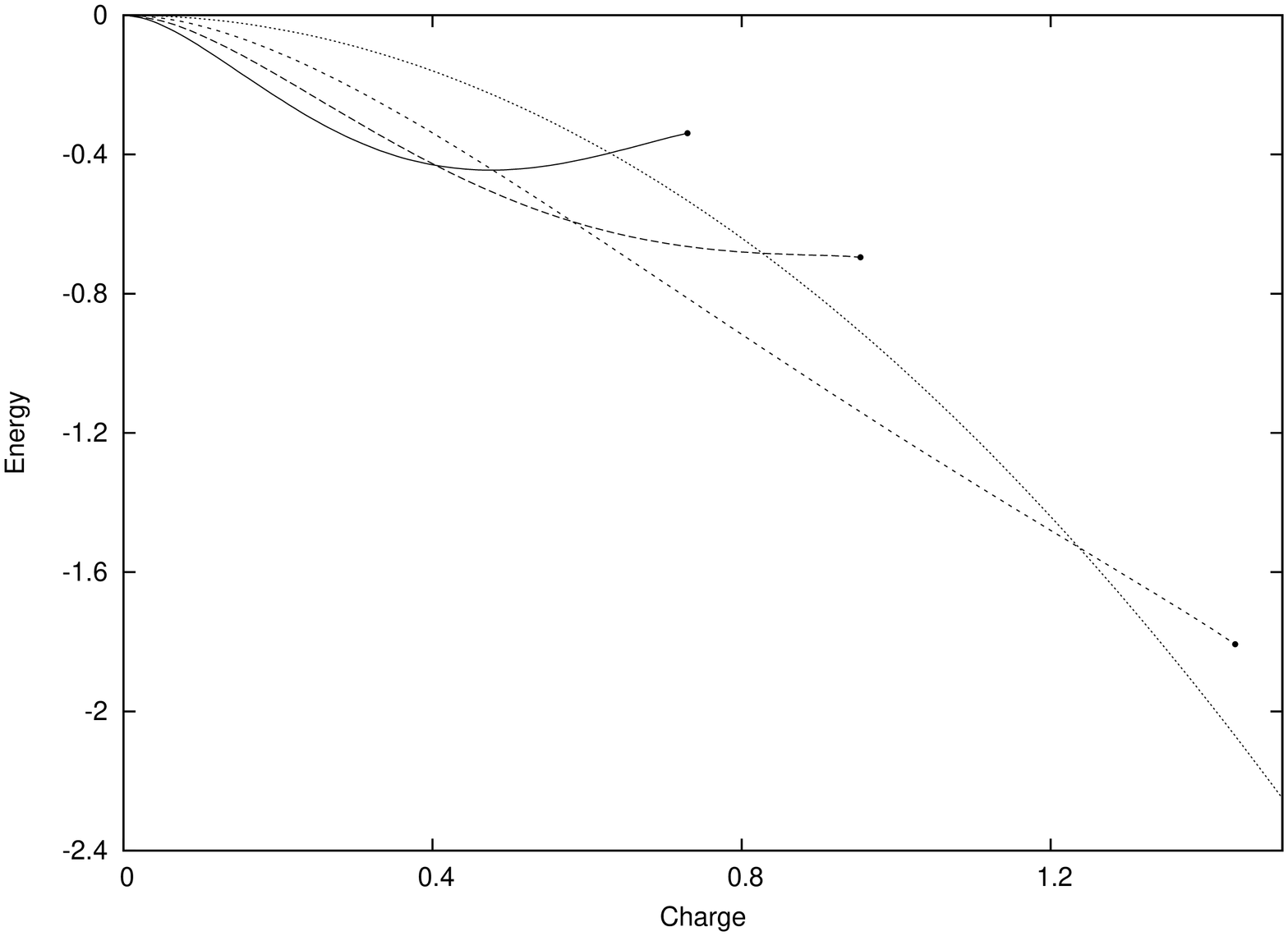}
\caption{Total energy in Ry of systems $(Z,e)$ (dotted line), $(2Z,e)$ at
  $R=R_{eq}$ (dashed line), $(3Z,e)$ at $R=R_{eq}$ (long-dashed line) and $(4Z,e)$ at $R=R_{eq}$ (solid line) as functions of the charge $Z$.
  $(Z,e)$ and $(2Z,e)$ curves cross at $Z=Z_{cross}=1.237$.
  Dashed curve ends at $Z=Z^{(1)}_{cr}=1.439$.
  Long-dashed curve ends at $Z=Z^{(2)}_{cr}=0.9539$.
  Dotted curve ends at $Z=Z^{(3)}_{cr}=0.736$
  }
\label{E:nz+e}
\end{figure}

\begin{figure}[ht]
  \centering
\psfrag{Equilibrium distance as function of the Charge}{}
\psfrag{Charge}{\huge{$Z$}}
\psfrag{Energy}{\huge{Total Energy}}
\psfrag{Zpe}{(Z,e)}
\psfrag{ZpZpe}{(Z,Z,e)}
\psfrag{ZpZpZpe}{(Z,Z,Z,e)}
\psfrag{ZpZpZpZpe}{(Z,Z,Z,Z,e)}
\includegraphics[angle=-90, width=1.0\textwidth]{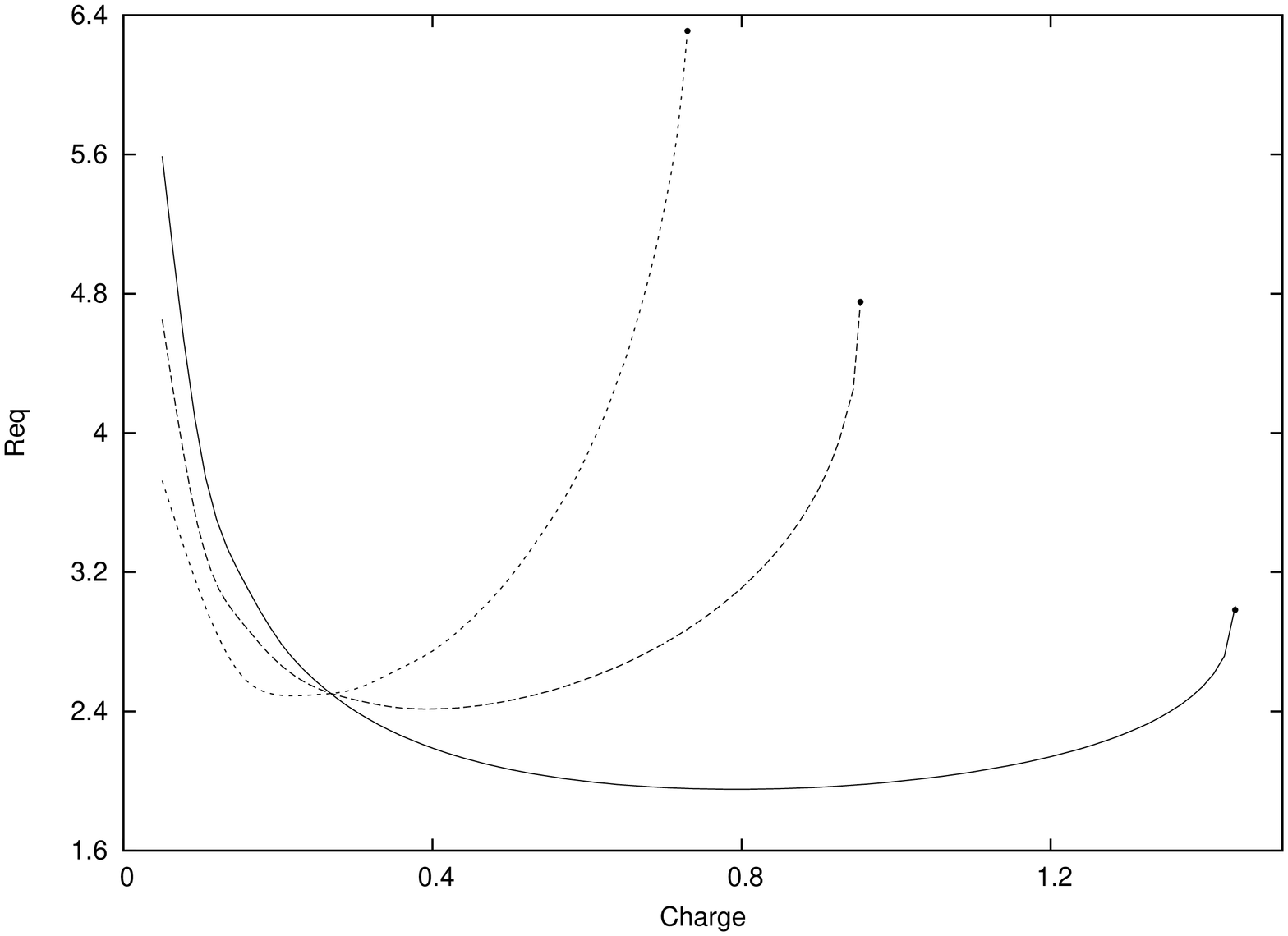}
\caption{Equilibrium distance in a.u. of systems $(2Z,e)$
  (dashed line), $(3Z,e)$ (long-dashed line) and $(4Z,e)$ (solid line) as functions of the charge $Z$. All curves cross at $Z=0.2977$ with
  $R_{eq}^{cross}=2.366$~a.u.
  Dashed curve ends at $Z=Z^{(1)}_{cr}=1.439$.
  Long-dashed curve ends at $Z=Z^{(2)}_{cr}=0.9539$.
  Dotted curve ends at $Z=Z^{(3)}_{cr}=0.736$.
  }
\label{R:nz+e}
\end{figure}

With such a six-parametric trial function (\ref{trial-2Ze})
\footnote{The Hamiltonian (\ref{Hnk}) at $k=1$ allows the separation of variables in elliptic coordinates (see e.g. \cite{LL}).
Every function $\psi_{1,2,3}$ admits a factorization in elliptic coordinates, although their linear superposition (\ref{trial-2Ze}), in general, does not. Factorization implies imposing a constraint on parameters $\al$'s. We do not impose such a constraint, however, for the parameters obtained as a result of the variational optimization the constraint is almost fulfilled. It indicates to a quality of the trial function (\ref{trial-2Ze}).}
the expected relative accuracy in total energy is $\approx 10^{-5}$, which is confirmed by an independent calculation based on use of highly accurate uniform approximation of the ground state eigenfunction \cite{TH:2011} (see a discussion below). The total energy $E(Z,R=R_{eq})$ is presented at Fig. \ref{E:nz+e} and the equilibrium distance is at Fig. \ref{R:nz+e}. Both curves are smooth without any indication to a special feature at the physical charge $Z=1$. At some charge (see below) the energy curves for $(2Z,e)$ and Z-atom $(Z,e)$ intersect. This crossing separates the domain of stability from metastability of the system $(2Z,e)$. The equilibrium distance {\it vs.} $Z$ is a smooth curve which has a minimum $R_{eq}=1.952$ a.u. at $Z=0.7924$, expectedly, with a decrease of $Z$ it grows to infinity. At critical charge $Z=1.439$ the equilibrium distance, where the potential curve $E=E(R)$ has the saddle point, is equal to 2.985 a.u.

It is interesting to study the approach of the total energy to the critical charge $Z \rar Z_{cr}$ from below. In order to do it we use the Puiseux expansion
\begin{equation}
\label{Pui}
    E_T(Z)= \sum_{n=0}^{\infty} a_n(Z_{cr}-Z)^{b_n}\ ,
\end{equation}
with the condition that $b_n < b_{n+1}$. Our goal is to find parameters $a_n$
and $b_n$ of this expansion. Restricting the expansion (\ref{Pui}) to a finite number of terms we make fit of the total energy calculated numerically,
see Table I.
\begin{table}[htb]
\centering
\begin{tabular}{|c|c|c|}
\hline
 \quad $Z$ \quad &\quad $E_T$ \quad &\quad Fit {}\quad \\
\hline\hline
 0.10 &\ -0.031019 \ &\ -0.03071\\ \hline
 0.15 &\ -0.064596 \ &\ -0.06455\\ \hline
 0.20 &\ -0.107149 \ &\ -0.10735\\ \hline
 0.25 &\ -0.157038 \ &\ -0.15725\\ \hline
 0.30 &\ -0.212917 \ &\ -0.21287\\ \hline
 0.35 &\ -0.273656 \ &\ -0.27336\\ \hline
 0.40 &\ -0.338292 \ &\ -0.33838\\ \hline
 1.30 &\ -1.614220 \ &\ -1.61422\
\\ \hline
 1.32 &\ -1.641112 \ &\ -1.64111\
\\ \hline
 1.34 &\ -1.668126 \ &\ -1.66813\
\\ \hline
 1.36 &\ -1.695327 \ &\ -1.69533\
\\ \hline
 1.38 &\ -1.722801 \ &\ -1.72280\
\\ \hline
 1.40 &\ -1.750671 \ &\ -1.75067\
\\ \hline
 1.41 &\ -1.764813 \ &\ -1.76481\
\\ \hline
 1.42 &\ -1.779144 \ &\ -1.77914\
\\ \hline
 1.43 &\ -1.793737 \ &\ -1.79373\
\\  \hline\hline
\end{tabular}
\caption{
Total energy $E_T$ of $(2Z,e)$ in Ry at equilibrium vs $Z$ obtained using (\ref{trial-2Ze}) and in the method \cite{TH:2011} compared to the result of the fit (\ref{fitenz2+}).
}
\label{fit_2Z}
\end{table}
The fit based on data from the domain $Z\in[1.30,1.43]$ (20 points) is:
\begin{eqnarray}
\label{fitenz2+}
 E_T(Z) &=& -1.8072 + 1.5538\ (Z_{cr} - Z) - 0.5719\ (Z_{cr} - Z)^{3/2} \\ \non
  &  +  & 0.1129(Z_{cr} - Z)^2+0.7777(Z_{cr}-Z)^{5/2}-0.4086(Z_{cr}-Z)^{5/2}+\ldots
\end{eqnarray}
where the critical point is
\begin{equation}
\label{Z2+cr}
    Z_{cr}^{(1)}\ =\ 1.439\ .
\end{equation}
This behavior indicates that critical point might be a square-root branch point.

There exists a charge for which a type of the binding of the system $(2Z,e)$
is changed from metastable, $(2Z,e) \to (Z,e)+Z$, to stable, $Z=Z_{cross}=1.237$ at
$R_{eq, cross}=2.184$ a.u. It corresponds to the crossing of two potential curves
on Fig. \ref{E:nz+e}. If $Z < Z_{cross}$ the system $(2Z,e)$ is stable, if $Z > Z_{cross}$ the system $(2Z,e)$ gets unstable, $(2Z,e) \to (Z,e)+Z$. Value of $Z_{cross}$ we calculated coincides with one found by Rebane \cite{rebane}.

As next we study the behavior of the total energy near the point of
crossing, $Z_{cross}$. From the left , $Z < Z_{cross}$, we find as the
result of the fit that the Puiseux expansion becomes the Taylor expansion
\[
   E_T = -1.5292 + 1.341\ (1.2366 - Z) + 0.08\ (1.2366 - Z)^2 + \ldots
\]
as well as from the right, $Z > Z_{cross}$, our data are also fit by
the Taylor expansion
\[
   E_T = -1.5292 + 1.340\ (Z - 1.2366) + 0.05\ (Z - 1.2366)^2 + \ldots
\]
Inside of the accuracy of data used these expansions do coincide. Therefore,
we do not see an indication to a branch point singularity contrary to the
statement in \cite{kais}. It is worth mentioning that the dependence of $R_{eq}$ on  $Z$ near $Z_{cross}$ is also very smooth, see Fig. \ref{R:nz+e}.

Another question to rise is a behavior of the total energy near point $Z=0$ which is the singular point of the Schr\"odinger equation.
Based on the fit of data from the domain $Z \in [0.1, 0.5]$ (six points, see Table I) we find that the Puiseux expansion (\ref{Pui}) becomes the Taylor expansion
\[
   E_T\ =\ -3.5258 Z^2 + 4.8922 Z^3 - 3.4121 Z^4 + \ldots \ .
\]
Such a behavior does not provide an indication to a singular nature of the point $Z=0$. However, the total energy can not be analytically continued to
$\mbox{Re} Z <0$.

\subsection{$(3Z,e)$}

\begin{figure}[ht]
\centering
\includegraphics[angle=00, width=0.40\textwidth]{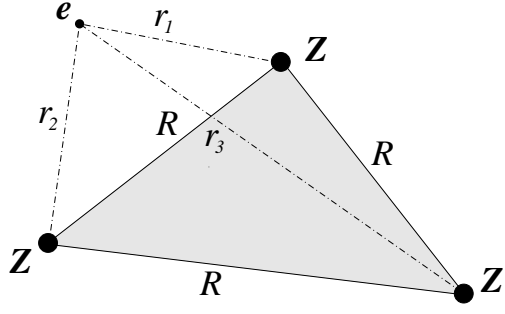}
\caption{Geometrical setting for $(3Z,e)$ system}
\label{z3++}
\end{figure}

Let us consider the electron in the electric field of three static charges $Z$:
$(3Z,e)$. In general, these charges form triangle, see Fig. \ref{z3++} as an
illustration. Such a system does not exist at $Z=1$ \cite{alijah}. Thus, there might exist a critical charge $Z <1$ for which the system gets bound, it separates the domain of the non-existence from existence of the bound state. Evidently, one of such critical charges is at $Z=0$, which is the singular point of the Schr\"odinger equation. Another one is at some $Z=Z_{cr} < 1$ (see \cite{alijah}). Calculations (see below) show that $Z_{cr}=0.9539$ with $R_{eq}=4.754$ a.u. Thus, the system $(3Z,e)$ exists for charges $0<Z<Z_{cr}$ always in a form of equilateral triangle, which is the optimal geometrical configuration. It was checked that this configuration is always stable with respect to small deviation.

In order to calculate the total energy $E(Z,R)$ the variational method is used.
We employ the physics-inspired trial functions \cite{turbinervar, turbinervar1, TL:2002} taking afterwards their linear superposition,
\begin{equation}
\label{trial-3Ze}
\Psi_{trial}= \sum_{j=1}^6 A_j \psi^{(j)}\ ,
\end{equation}
here $A_j$ are linear parameters. Each function $\psi^{(j)}$ is
chosen in such a way to describe a certain physical situation of the system.
In general, $\psi^{(j)}$ has the form of a symmetrized product of three
$1s$-Coulomb orbitals (Slater functions)
\begin{equation}
\psi_C = e^{-\al_{1}r_1-\al_{2}r_2-\al_{3}r_3}\ .
\end{equation}
Let us give a brief description of each of them \cite{alijah}:

\begin{description}
\item[$\psi^{(1)}$:] All $\al$'s are chosen to be equal to $\al_1$,
\begin{equation}
  \psi^{(1)} =
  e^{-\al_{1}(r_1+r_2+r_3)}\,.
  \end{equation}

It is a Heitler-London type function. This
corresponds to \emph{coherent} interaction between the electron and
all centers. Supposedly, it describes the system at small intercenter
distances and, probably, the equilibrium configuration. It is
verified {\it a posteriori}.

\item[$\psi^{(2)}$:] Two $\al$'s are equal to zero and the remaining
  one is set to be equal to $\al_2$,
\begin{equation}
  \psi^{(2)}
  =e^{-\alpha_{2}r_1}+e^{-\al_{2}r_2}+e^{-\al_{2}r_3} \ .
\end{equation}

It is a Hund-Mulliken type function.  This
function possibly describes the system at large distances, where
essentially the electron interacts with only one center at a time thus
realizing \emph{totally incoherent} interaction.

\item[$\psi^{(3)}$:] One $\al$ is equal to zero, two others are
  different from zero but equal to each other and to $\al_3$,
 \begin{equation}
 \label{psi3}
  \psi^{(3)}
  =  e^{-\al_{3}(r_1+r_2)} + e^{-\al_{3}(r_1+r_3)}
           + e^{-\al_{3}(r_2+r_3)}\ .
\end{equation}

It is assumed that this function describes the system $(2Z,e)$
plus center when a triangle is of a sufficiently small size. In
fact, it is the Heitler-London function of $(2Z,e)$-system symmetrized
over centers.

\item[$\psi^{(4)}$:] One $\al$ is equal to zero and two others are
  different from each other being equal to $\al_{4,5}$, respectively,
\begin{eqnarray}
\label{psi4}
   \psi^{(4)}
   &=&  e^{-\al_{4}r_1-\al_{5}r_2} +
   e^{-\al_{4}r_2-\al_{5}r_1} +
   e^{-\al_{4}r_1-\al_{5}r_3} \non \\
   &+&   e^{-\al_{4}r_3-\al_{5}r_1} +
   e^{-\al_{4}r_2-\al_{5}r_3} + e^{-\al_{4}r_3-\al_{5}r_2}\ .
\end{eqnarray}

It is assumed that this function describes the system $(2Z,e)$ plus
one center. In fact, it is the Guillemin-Zener
function of the $(2Z,e)$-system then symmetrized over centers.  If $\al_4=\al_5$, the function $\psi^{(4)}$ is reduced to $\psi^{(3)}$. If $\al_4=0$, the function $\psi^{(4)}$ is reduced to $\psi^{(2)}$. Hence,
$\psi^{(4)}$ is a non-linear interpolation between $\psi^{(2)}$ and
$\psi^{(3)}$.  It has to describe intermediate intercenter distances.

\item[$\psi^{(5)}$:] Two $\al$'s are equal but the third one is
    different,
 \begin{eqnarray}
   \psi^{(5)}
   &=&  e^{-\alpha_{6}r_1-\alpha_{6}r_2-\alpha_{7}r_3}
   + e^{-\alpha_{6}r_1-\alpha_{6}r_3-\alpha_{7}r_2}   \non\\
   &+&  e^{-\alpha_{6}r_2-\alpha_{6}r_1-\alpha_{7}r_3}
   + e^{-\alpha_{6}r_2-\alpha_{6}r_3-\alpha_{7}r_1}   \non\\
   &+&  e^{-\alpha_{6}r_3-\alpha_{6}r_1-\alpha_{7}r_2}
   + e^{-\alpha_{6}r_3-\alpha_{6}r_2-\alpha_{7}r_1}\ .
  \end{eqnarray}

It describes a ``mixed'' state of three Z-hydrogen atoms. If
$\al_6=\al_7$, the function $\psi^{(5)}$ is reduced to
$\psi^{(1)}$. If $\al_6=0$, the function $\psi^{(5)}$ is reduced
to $\psi^{(2)}$. If $\al_7=0$, the function $\psi^{(5)}$ is
reduced to $\psi^{(3)}$.  Hence, $\psi^{(5)}$ is a non-linear
interpolation between $\psi^{(1)}$, $\psi^{(2)}$ and
$\psi^{(3)}$. As function~(\ref{psi4}) this is a type of
Guillemin-Zener function and should describe intermediate
intercenter distances.

\item[$\psi^{(6)}$:] All $\al$'s are different,
\begin{eqnarray}
\label{psi6}
  \psi^{(6)}
  & = & e^{-\alpha_{8}r_1-\alpha_{9}r_2-\alpha_{10}r_3}
  + e^{-\alpha_{8}r_1-\alpha_{9}r_3-\alpha_{10}r_2} \non\\
  & + & e^{-\alpha_{8}r_2-\alpha_{9}r_1-\alpha_{10}r_3}
  + e^{-\alpha_{8}r_2-\alpha_{9}r_3-\alpha_{10}r_1} \non\\
  & + & e^{-\alpha_{8}r_3-\alpha_{9}r_1-\alpha_{10}r_2}
  + e^{-\alpha_{8}r_3-\alpha_{9}r_2-\alpha_{10}r_1}\ .
\end{eqnarray}

This is a general non-linear interpolation of all functions $\psi^{(1-5)}$.
\end{description}

The total number of parameters of the function (\ref{trial-3Ze}) is equal
to 15, where five are linear ones.  Note that without a loss of
generality the parameter $A_6$ in (\ref{trial-3Ze}) can be fixed,
putting $A_6 = 1$. We expect this function provides a relative accuracy
$\sim 10^{-3}$ in total energy.

As a result of variational study for fixed $Z$ the optimal geometric configuration is always the equilateral triangle. It was checked that this configuration is always stable with respect to small deviation. On Fig. \ref{E:nz+e} the total energy dependence for $(3Z,e)$ at the equilibrium configuration is given.  It is a smooth monotonous curve which ends at $Z=Z_{cr} < 1$. At some charges this curve intersects with the energy curves for $(2Z,e)$ and Z-atom, $(Z,e)$. These crossings separate domains of stability from different domains of metastability of the system (see below). On Fig. \ref{R:nz+e} the equilibrium distance between nearest static charges (the side of the equilateral triangle) is shown. It is a smooth curve which has a minimum $R^{min}_{eq}=2.413$ a.u. at $Z=0.391$ and it grows to infinity with a decrease of $Z$. At critical charge $Z=0.9539$ the equilibrium distance, where the potential curve $E=E(R)$ has the saddle point, is equal to 4.754 a.u. It is worth mentioning that the $R_{eq}$ curves for $(3Z,e)$ and $(2Z,e)$ intersect at $Z=0.2670 $ with $R_{eq}=2.506$ a.u.

It is interesting to study the approach of the total energy to the critical charge $Z \rar Z_{cr}$ from below. In order to do it we use a general Puiseux expansion (\ref{Pui}). Eventually, the behavior of the total energy close to critical charge $Z_{cr}$, as a result of the fit, is given by the following terminated Puiseux expansion:
\begin{eqnarray}
\label{z3++enfit}
E(Z)  & = & -0.6954+0.2700(Z_{cr}-Z)-1.0357(Z_{cr}-Z)^{3/2} - 1.3360(Z_{cr}-Z)^2
\\ \non
      &  & -0.1350(Z_{cr}-Z)^{5/2}+2.3395(Z_{cr}-Z)^3
           -1.8714(Z_{cr}-Z)^{7/2} \ ,
\end{eqnarray}
where the critical point is
\begin{equation}
\label{Z3++cr}
    Z_{cr}^{(2)}\ =\ 0.9539\ .
\end{equation}
The fit (\ref{z3++enfit}) is based on data from the domain $Z \in [0.80,0.93]$ (19 points, see Table II). This behavior indicates that the critical point might be a square-root branch point.

\begin{table}[htb]
\centering
\begin{tabular}{|c|c|c|}
\hline
 $Z$\qquad &\ $E_T$\quad &\quad Fit \quad \\
\hline\hline
 0.10 &       -0.057230&     -0.056880   \\\hline
 0.15 &       -0.111367&     -0.111714   \\\hline
 0.20 &       -0.173248&	 -0.173581   \\\hline
 0.25 &       -0.238656&	 -0.238205   \\\hline
 0.30 &       -0.304100&	 -0.304235    \\
\hline\hline
 0.80 &       -0.680137&     -0.680137    \\\hline
 0.82 &       -0.682953&     -0.682953    \\\hline
 0.84 &       -0.685194&     -0.685194    \\\hline
 0.86 &       -0.686969&     -0.686969    \\\hline
 0.88 &       -0.688417&     -0.688417    \\\hline
 0.90 &       -0.689716&     -0.689716    \\\hline
 0.91 &       -0.690385&     -0.690385    \\\hline
 0.92 &       -0.691123&     -0.691123    \\\hline
 0.93 &       -0.691991&     -0.691991    \\\hline
\end{tabular}
\caption{Total energy $E_T$ of $(3Z,e)$ in Ry at equilibrium vs $Z$ obtained using (\ref{trial-3Ze}) compared to the result of the fit (\ref{z3++enfit}).
}
\label{fit_3Z}
\end{table}

There are two points of crossing for the energy curve $(3Z,e)$ at Fig. \ref{E:nz+e}.  The first one is $Z_{cross}^{(1)}=0.8269$ with $R_{eq}=3.234$ a.u. for the crossing of $(3Z,e)$ and the $(Z,e)$. The second one is  $Z_{cross}^{(2)}=0.5811$ with $R_{eq}=2.640$ a.u. for the crossing of $(3Z,e)$ and the $(2Z,e)$ at $R_{eq}= 2.008$ a.u.
For charges $Z \in (0.8269,0.9537)$ for the triangular equilateral
configuration the system is metastable with two decay channels
\begin{eqnarray}
(3Z,e) \to (Z,e)+ Z + Z \ ,  \non \\
(3Z,e) \to {\rm Z}_2^{+} + Z\ , \non
\end{eqnarray}
while for $Z\in(0.5811,0.8269)$ system is metastable with single decay channel
\[
(3Z,e) \to (2Z,e) + Z\ ,
\]
and, finally, for $Z<0.5811$ the system is stable. A study of the Puiseux expansions near $Z_{cross}^{(1)}$ as well as $Z_{cross}^{(2)}$ from above and below show that they are the Taylor expansions which do coincide within the accuracy of data used and the obtained parameters of the fit. They do not give an indication that these points are branch points. It also is worth mentioning that the dependence of $R_{eq}$ on  $Z$ near $Z_{cross}^{(1,2)}$ is also very smooth, see Fig. \ref{R:nz+e}.

Another question to rise is a behavior of the total energy near the critical point at $Z=0$ which is the singular point of the Schr\"odinger equation. Based on the fit of data from the domain $Z \in [0.1, 0.7]$ (seven points, see Table II) we find that the Puiseux expansion becomes the Taylor expansion
\[
   E_T\ =\ -7.4257 Z^2 + 19.3244 Z^3 - 19.4662 Z^4 + \ldots \ .
\]
Such a behavior does not provide an indication to singular nature of the point $Z=0$. However, the total energy can not be analytically continued to
$\mbox{Re} Z <0$.

\subsection{$(4Z,e)$}

\begin{figure}[htb!]
\centering
\includegraphics[angle=00, width=0.5\textwidth]{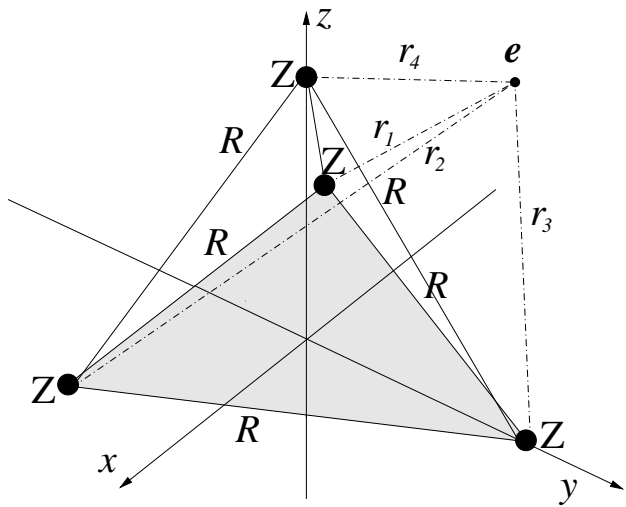}
\caption{Geometrical setting for $(4Z,e)$ system}
\label{tetra1}
\end{figure}

This Coulomb system consists of four static $Z$-charges and one electron,
$(4Z,e)$. It is worth anticipating that the most symmetrical configuration where $Z\!-$charges are placed on the vertexes of a tetrahedron, see
Fig.~\ref{tetra1}, is optimal. It was checked that this configuration is always stable with respect to small deviations. It is certain that for $Z=1$ there is a bound state, the system H$_4^{3+}$ does not exist.

{\it Trial Functions.}
The variational method was used to obtain all numerical results. Trial
function is taken in a form of linear superposition of three functions:
\begin{equation}
\label{trial-4Ze}
\Psi_{trial}= \sum_{j=1}^3 A_j \psi^{(j)}\ ,
\end{equation}
where $A_j$ are linear parameters. Each function $\psi^{(j)}$ is
chosen in such a way as to describe different physical characteristics
of the system \cite{turbinervar, turbinervar1}. In general,
$\psi^{(j)}$ has the form of a symmetrized product of four
$1s$-Coulomb orbitals (Slater functions)
\begin{equation}
\psi_C = e^{-\al_{1}r_1-\al_{2}r_2-\al_{3}r_3-\al_4r_4}\ .
\end{equation}
Let us give a brief description of each of them:
\begin{description}

\item[$\psi^{(1)}$]: All $\al$'s are chosen to equal to $\al_1$,
\begin{equation}
  \psi^{(1)} =  e^{-\alpha_{1}(r_1+r_2+r_3+r_4)}\,.
  \end{equation}
  It is a Heitler-London type function. This
  corresponds to \emph{coherent} interaction between the electron and
  all protons. Supposedly, it describes the system at small interproton
  distances and, probably, the equilibrium configuration.

\item[$\psi^{(2)}$]: Three $\al$'s are equal to zero and the
  remaining one is set to be equal to $\al_2$,
\begin{equation}
  \psi^{(2)}
  =e^{-\al_{2}r_1}+e^{-\al_{2}r_2}+e^{-\al_{2}r_3}+e^{-\al_{2}r_4} \ .
\end{equation}
It is a Hund-Mulliken type function.
    This function possibly describes the system at large
    distances, where essentially the electron interacts with only one
    proton at a time thus realizing \emph{incoherent} interaction.

\item[$\psi^{(3)}$]: All $\al$'s are different from each other, and
different from zero.
\begin{eqnarray}
\psi^{(3)}&=&e^{-\al_{3} r_{1}-\al_{4} r_{2}-\al_{5} r_{3}-\al_{6} r_{4}}+e^{-\alpha_{3} r_{1}-\al_{4} r_{2}-\al_{6} r_{3}-\al_{5} r_{4}}\nonumber\\
&+&e^{-\alpha_{3} r_{1}-\alpha_{5} r_{2}-\alpha_{4} r_{3}-\al_{6} r_{4}}+e^{-\alpha_{3} r_{1}-\alpha_{5} r_{2}-\alpha_{6} r_{3}-\al_{4} r_{4}}\nonumber\\
&+&e^{-\alpha_{3} r_{1}-\alpha_{6} r_{2}-\alpha_{4} r_{3}-\al_{5} r_{4}}+e^{-\alpha_{3} r_{1}-\alpha_{6} r_{2}-\alpha_{5} r_{3}-\al_{4} r_{4}}\nonumber\\
&+&e^{-\alpha_{4} r_{1}-\alpha_{3} r_{2}-\alpha_{5} r_{3}-\alpha_{6} r_{4}}+e^{-\alpha_{4} r_{1}-\alpha_{3} r_{2}-\alpha_{6} r_{3}-\alpha_{5} r_{4}}\nonumber\\
&+&e^{-\alpha_{4} r_{1}-\alpha_{5} r_{2}-\alpha_{3} r_{3}-\alpha_{6} r_{4}}+e^{-\alpha_{4} r_{1}-\alpha_{5} r_{2}-\alpha_{6} r_{3}-\alpha_{3} r_{4}}\nonumber\\
&+&e^{-\alpha_{4} r_{1}-\alpha_{6} r_{2}-\alpha_{3} r_{3}-\alpha_{5} r_{4}}+e^{-\alpha_{4} r_{1}-\alpha_{6} r_{2}-\alpha_{5} r_{3}-\alpha_{3} r_{4}}\nonumber\\
&+&e^{-\alpha_{5} r_{1}-\alpha_{3} r_{2}-\alpha_{4} r_{3}-\alpha_{6} r_{4}}+e^{-\alpha_{5} r_{1}-\alpha_{3} r_{2}-\alpha_{6} r_{3}-\alpha_{4} r_{4}}\nonumber\\
&+&e^{-\alpha_{5} r_{1}-\alpha_{4} r_{2}-\alpha_{3} r_{3}-\alpha_{6} r_{4}}+e^{-\alpha_{5} r_{1}-\alpha_{4} r_{2}-\alpha_{6} r_{3}-\alpha_{3} r_{4}}\nonumber\\
&+&e^{-\alpha_{5} r_{1}-\alpha_{6} r_{2}-\alpha_{3} r_{3}-\alpha_{4} r_{4}}+e^{-\alpha_{5} r_{1}-\alpha_{6} r_{2}-\alpha_{4} r_{3}-\alpha_{3} r_{4}}\nonumber\\
&+&e^{-\alpha_{6} r_{1}-\alpha_{3} r_{2}-\alpha_{4} r_{3}-\alpha_{5} r_{4}}+e^{-\alpha_{6} r_{1}-\alpha_{3} r_{2}-\alpha_{5} r_{3}-\alpha_{4} r_{4}}\nonumber\\
&+&e^{-\alpha_{6} r_{1}-\alpha_{4} r_{2}-\alpha_{3} r_{3}-\alpha_{5} r_{4}}+e^{-\alpha_{6} r_{1}-\alpha_{4} r_{2}-\alpha_{5} r_{3}-\alpha_{3} r_{4}}\nonumber\\
&+&e^{-\alpha_{6} r_{1}-\alpha_{5} r_{2}-\alpha_{3} r_{3}-\alpha_{4} r_{4}}+e^{-\alpha_{6} r_{1}-\alpha_{5} r_{2}-\alpha_{4} r_{3}-\alpha_{3} r_{4}}
\end{eqnarray}

\end{description}
We can see that trial functions $\psi^{(1)}$ and $\psi^{(2)}$ are
particular cases of the general trial function $\psi^{(3)}$.

There might exist two critical charges which separates the domain of
existence from non-existence of bound states. One such a critical
charge is at $Z=0$. Another one is at some $Z=Z_{cr} < 1$. Calculations
(see below) show that $Z_{cr}=0.736$ at $R_{eq}=6.50$ a.u. , where the potential curve $E=E(R)$ has the saddle point. Thus, the system
$(4Z,e)$ can exist for charges $0<Z<Z_{cr}$. The energy dependence at equilibrium distance $R_{eq}$ is a smooth function, see Fig. \ref{E:nz+e}.
The optimal geometric configuration is always the tetrahedron. It was checked that this configuration is always stable with respect to small deviations. It is a smooth monotonous curve which ends at $Z=Z_{cr} < 1$. At some charges it intersects the energy curves for $(3Z,e)$, $(2Z,e)$ and Z-atom $(Z,e)$. These crossings separate domains of stability from metastability of the system (see below). On Fig. \ref{R:nz+e} the equilibrium distance between nearest static charges (the side of the tetrahedron) is shown. It is a smooth curve which has a minimum $R_{eq}=2.485$ a.u. at $Z=0.2218$ and it grows to infinity with a decrease of $Z$. It is quite amusing that all three equilibrium distance curves for $(4Z,e)$, $(3Z,e)$ and $(2Z,e)$ intersect for $Z=0.2670$ with $R_{eq}=2.506$ a.u.

Behavior of the energy as function of the charge close to critical
charge $Z_{cr}$ is given by the terminated Puiseux expansion:
\begin{eqnarray}
\label{z4+++enfit}
E(Z)& = &-0.3368-0.2793(Z_{cr}-Z)-1.5995(Z_{cr}-Z)^{3/2}\\ \non
    & + &2.0214(Z_{cr}-Z)^2+0.9224(Z_{cr}-Z)^{5/2}\ +\ \ldots,
\end{eqnarray}
where the critical point is
\begin{equation}
\label{Z4+++cr}
    Z_{cr}^{(3)}\ =\ 0.736 \ .
\end{equation}
The fit (\ref{z4+++enfit}) is based on data from the domain $Z\in[0.60,0.72]$
(12 points, see Table III). This behavior indicates that critical point might be a
square-root branch point.

\begin{table}[htb]
\centering
\begin{tabular}{|c|c|c|}
\hline
 $Z$ & $E_T$ & Fit \\
\hline\hline
 0.10&     -0.092030&	    -0.092063           \\\hline
 0.11&	   -0.106530&	    -0.106491           \\\hline
 0.12&	   -0.121220&	    -0.121190           \\\hline
 0.13&	   -0.136020&	    -0.136072           \\\hline
 0.14&	   -0.151080&	    -0.151070           \\\hline
 0.15&	   -0.166150&	    -0.166145           \\\hline
\hline
 0.60&         -0.411356&     -0.411387            \\\hline
 0.62&         -0.401055&     -0.401025            \\\hline
 0.64&         -0.389997&     -0.389987            \\\hline
 0.66&         -0.378426&     -0.378456            \\\hline
 0.68&         -0.366652&     -0.366675            \\\hline
 0.70&         -0.355022&     -0.354995            \\\hline
 0.72&         -0.344089&     -0.344021            \\\hline
\end{tabular}
\caption{Total energy $E_T$ of $(4Z,e)$ in Ry at equilibrium vs $Z$ obtained variationally using the trial function (\ref{trial-4Ze}) compared to the result of the fit (\ref{z4+++enfit}).
}
\label{fit_4Z}
\end{table}

There are three points of crossing for the energy curve $(4Z,e)$ at Fig. \ref{E:nz+e}.  The first one is for the crossing of $(4Z,e)$ and the $(Z,e)$ energy curve at $Z_{cross}^{(1)}=0.6290$ with $R_{eq}=4.187$ a.u. The second one is at $Z_{cross}^{(2)}=0.4798$ with $R_{eq}=3.08$ a.u. for the crossing of $(4Z,e)$ and the $(2Z,e)$ energy curve at $R_{eq}= 2.086$ a.u. The third one is  $Z_{cross}^{(3)}=0.4065$ with $R_{eq}=2.83$ a.u. for the crossing of $(4Z,e)$ and the $(3Z,e)$ energy curve at $R_{eq}= 2.413$ a.u.

For charges $Z\in (0.6290,0.7360)$ for the triangular equilateral
configuration the system is metastable with three decay channels
\begin{eqnarray}
(4Z,e) \to (Z,e) + Z+Z+Z \non \\
(4Z,e) \to (2Z,e) + Z+Z \non\\
(4Z,e) \to (3Z,e) + Z
\end{eqnarray}
For $Z \in (0.4798,0.6290)$ the system is metastable with two decay channels
\begin{eqnarray}
(4Z,e) \to (2Z,e) + Z+Z \quad ,\non \\
(4Z,e) \to (3Z,e) + Z
\end{eqnarray}
For $Z\in(0.4065,0.4798)$ system is metastable with one decay channel
\begin{equation}
(4Z,e) \to (3Z,e) + Z
\end{equation}
and, finally, for $0 < Z < 0.4065$ the system gets stable.

A question to rise is about behavior of the total energy near the critical point at $Z=0$ which is the singular point of the Schr\"odinger equation. Based on the fit of data from the domain $Z \in [0., 0.15]$ (seven points, see Table III) we find that the Puiseux expansion becomes the Taylor expansion
\[
   E_T\ =\ -14.3871 Z^2 + 62.0529 Z^3 - 102.4490 Z^4 + \ldots \ .
\]
Such a behavior does not provide an indication to singular nature of the point $Z=0$. However, the total energy can not be analytically continued to
$\mbox{Re} Z <0$.


\section{Two-electron molecular systems}

\subsection{$(2Z,e,e)$}

The system $(2Z,e,e)$ consists of two charged centers $Z$ and two electrons. For $Z=1$ it is the celebrated H$_2$ molecule when for $Z=2$ it is the Helium molecular ion He$_2^{(++)}$ which is metastable system. It is obvious that for large $Z$ the system is unbound as well as for negative $Z$. Thus, there are two singular points: $Z=0$ where the potential "changes" sign and $Z_{cr}>2$ which is a critical point separating the domain of the existence from the domain of non-existence of the solutions in the Hilbert space. It seems natural that the ground state when exists is the spin-singlet state. Calculations (see below) show that the critical charge $Z_{cr}=2.250$ at $R_{eq}=1.532$ a.u. Thus, the system $(2Z,e,e)$ exists for charge $0<Z<Z_{cr}$.

\begin{figure}[ht]
\includegraphics[angle=00, width=0.30\textwidth]{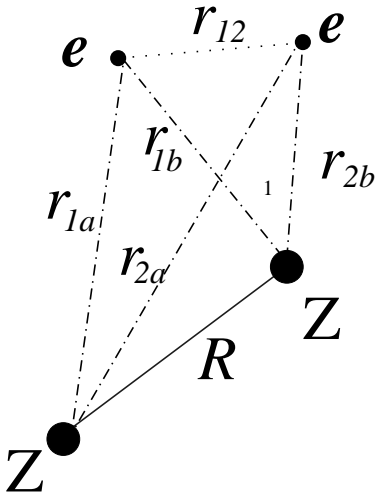}
\caption{Geometrical setting for $(2Z,e,e)$ system}
\end{figure}

{\it Trial Functions.}
To calculate the total energy of the $(2Z,e,e)$ system as a function of charge $Z$ the variational method is used. Exponential correlated trial functions with proper symmetrization are employed as well as their linear superposition. In general, the basic trial function $\psi^{(j)}$ is taken in the form of symmetrized product of four $1s$-Coulomb orbitals (Slater functions) and correlation function in exponential form \cite{TG:2007},
\begin{eqnarray}
\label{trial-2Z2e}
\psi_g&=&e^{-\al_{1} r_{1a}-\al_{2} r_{1b}-\al_{3} r_{2a}-\al_{4} r_{2b}+\gamma r_{12}}\non\\
&+&e^{-\al_{3} r_{1a}-\al_{4} r_{1b}-\al_{1} r_{2a}-\al_{2} r_{2b}+\gamma r_{12}}\non\\
&+&e^{-\al_{2} r_{1a}-\al_{1} r_{1b}-\al_{4} r_{2a}-\al_{3} r_{2b}+\gamma r_{12}}\non\\
&+&e^{-\al_{2} r_{1a}-\al_{1} r_{1b}-\al_{4} r_{2a}-\al_{3} r_{2b}+\gamma r_{12}}
\ .
\end{eqnarray}
Recently, it was shown that a linear superposition of three functions (\ref{trial-2Z2e}) leads to the most accurate ground state energy for the H$_2$-molecule among a few parametric trial functions.

\begin{figure}[ht]
\centering
\psfrag{Charge}{\huge{$Z$}}
\psfrag{Energy}{\huge{Total energy}}
\includegraphics[angle=-90, width=1.0\textwidth]{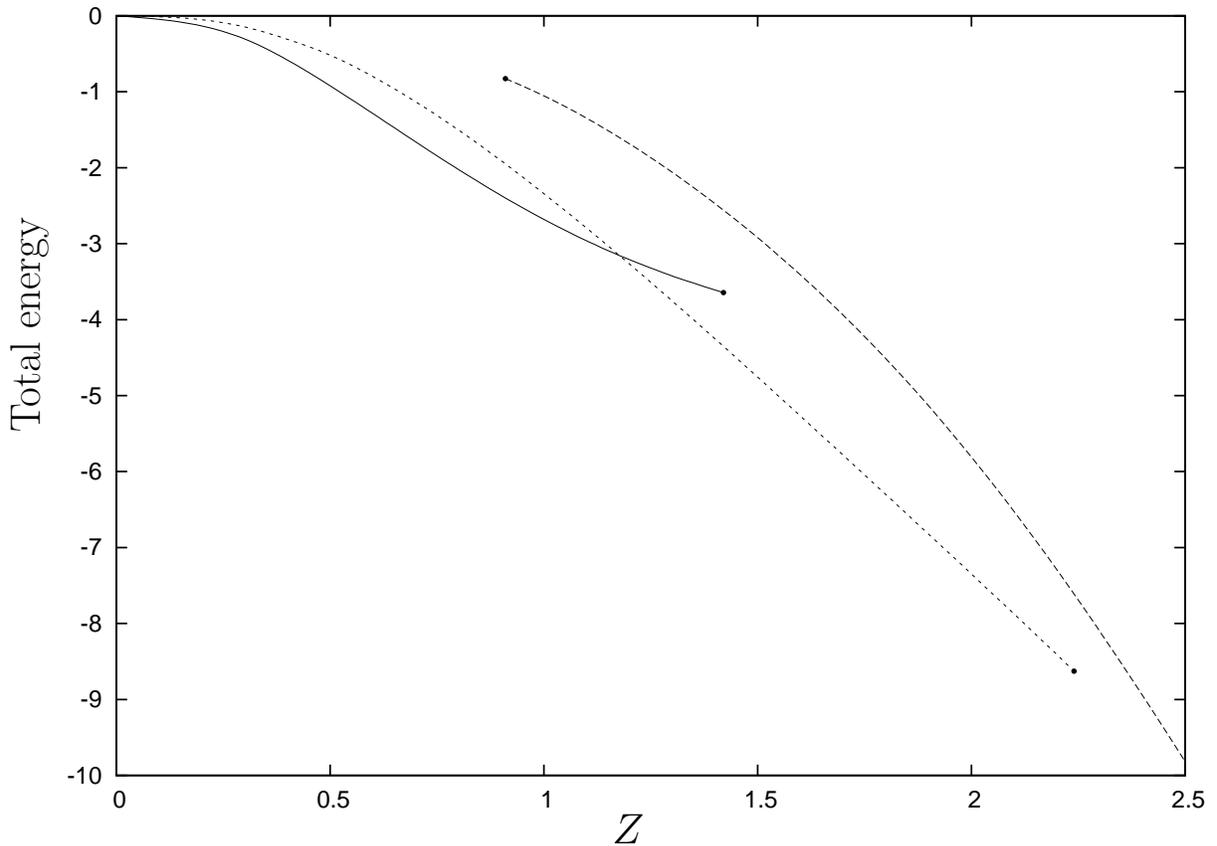}
\caption{Total energy $E_T$ in Ry vs $Z$ for two-electron systems
         in equilibrium at $R=R_{eq}$:
         $(2Z,e,e)$  (solid line) and $(3Z,e,e)$ (dashed line), and for comparison
         for $(Z,e,e)$  (long-dashed line).
         Both curves intersect at $Z=1.1767$.
         Solid line ends at $Z=Z^{(1)}_{cr}=2.250$.
         Dashed line ends at $Z=Z^{(2)}_{cr}=1.433$.
     }
\label{E:nz+ee}
\end{figure}

\begin{figure}[ht]
  \centering
\psfrag{Equilibrium distance as function of the Charge}{}
\psfrag{Charge}{\huge{$Z$}}
\psfrag{Energy}{\huge{Total Energy}}
\psfrag{Zpe}{(Z,e)}
\psfrag{ZpZpe}{(Z,Z,e)}
\psfrag{ZpZpZpe}{(Z,Z,Z,e)}
\psfrag{ZpZpZpZpe}{(Z,Z,Z,Z,e)}
\includegraphics[angle=-90, width=1.0\textwidth]{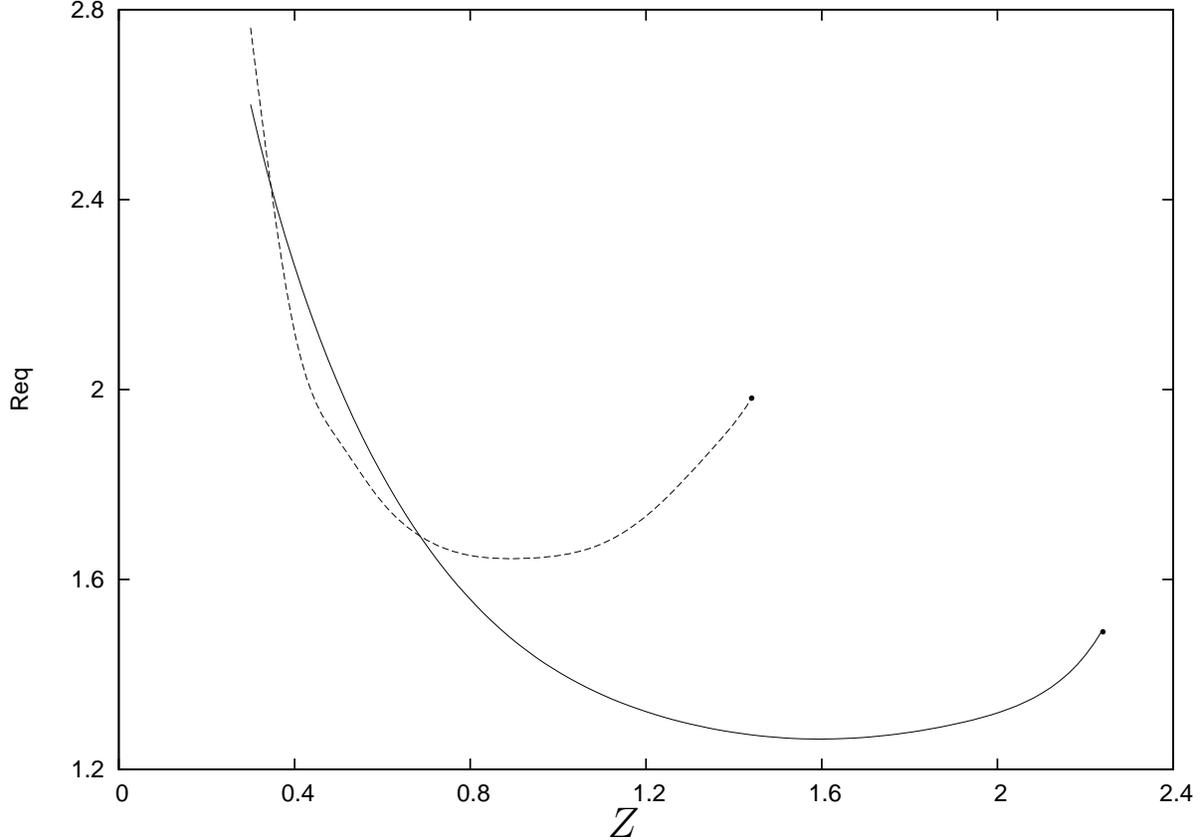}
\caption{Equilibrium distance $R_{eq}$ in a.u. of system $(2Z,e,e)$
  (solid line) and $(3Z,e,e)$ (dashed line) as functions of the charge $Z$.
  Curves cross twice at $Z=0.3460$ and at $Z=0.6851$.
  Solid line ends at $Z=Z^{(1)}_{cr}=2.250$.
  Dashed line ends at $Z=Z^{(2)}_{cr}=1.433$.
  }
\label{R:nz+ee}
\end{figure}

With such a function the expected relative accuracy in total energy is $\approx 10^{-3}$. The total energy $E(Z,R=R_{eq})$ is presented at Fig. \ref{E:nz+ee} and the equilibrium distance is at Fig. \ref{R:nz+ee}. Both curves are smooth without any indication to a special feature at the physical charge $Z=1,2$. At charge $Z=1.1767$ the energy curves for $(2Z,e,e)$ and $(3Z,e,e)$ intersect. The equilibrium distance is a smooth curve which has a minimum $R_{eq}=1.264$ a.u. at $Z=1.596$ and it grows to infinity with a decrease of $Z$. At critical charge $Z=2.250$ the equilibrium distance, where the potential curve $E=E(R)$ has the saddle point, is equal to 1.532 a.u.

Behavior of the energy as function of the charge close to critical charge
$Z<Z_{cr}$ is given by the terminated Puiseux expansion (see (\ref{Pui})):
\begin{eqnarray}
\label{z2enfit}
E(Z) & = & -8.6835 + 5.5238(Z_{cr}-Z) - 0.2982(Z_{cr}-Z)^{3/2}\\ \non
     & - &  0.3166(Z_{cr}-Z)^2 + 0.3577(Z_{cr}-Z)^{5/2} + \ldots \ ,
\end{eqnarray}
where the critical point is
\begin{equation}
\label{Z2cr}
    Z_{cr}=2.250 \ .
\end{equation}
The fit (\ref{z2enfit}) is based on data from the domain $Z\in[1.80,2.22]$
(12 points, see Table IV). This behavior indicates that critical point might be a square-root branch point.

\begin{table}[htb]
\centering
\begin{tabular}{|c|c|c|}
\hline
 $Z$\ &\ $E_T$\ &\ Fit\ \\
\hline\hline
 0.30   & -0.1464&-0.1496 \\\hline
 0.40   & -0.3091&-0.3037 \\\hline
 0.50   & -0.5177&-0.5211 \\\hline
 0.60   & -0.8001&-0.7993 \\\hline
\hline
 1.80   & -6.30334 & -6.30331 \\\hline
 1.90   & -6.82475 & -6.82475 \\\hline
 2.00   & -7.34768 & -7.34841 \\\hline
 2.10   & -7.87639 & -7.87623 \\\hline
 2.20   & -8.41058 & -8.41121 \\\hline
 2.22   & -8.52070 & -8.51953 \\\hline
\end{tabular}
\caption{Total energy $E_T$ of $(2Z,e,e)$ in Ry at equilibrium distance vs $Z$ obtained using the trial function (\ref{trial-2Z2e}) compared to the result of the fit (\ref{z2enfit}).
}
\label{fit_2Z+2e}
\end{table}

There are two points of crossing for the energy curve $(2Z,e,e)$ displayed at Fig. \ref{E:nz+ee}.  The first one is at $Z_{cross}^{(1)}=1.7026$ with $R_{eq}=1.268$ a.u. for the crossing of $(2Z,e,e)$ and two atoms $(Z,e)$. The second one is  $Z_{cross}^{(2)}=0.4501$ with $R_{eq}=2.126$ a.u. for the crossing of $(2Z,e,e)$ and the $(2Z,e)$ at $R_{eq}= 2.119$ a.u. and two atoms $(Z,e)$.

For $Z \in (1.7026, 2.250)$ the system $(2Z,e,e)$ is metastable, there is decay channel
\[
  (2Z,e,e) \rar (Z,e) + (Z,e)\ ,
\]
while for $Z \in (0.4501,1.7026)$ system is stable and
for $Z < Z_{cross}^{(2)}=0.4501$ it seemingly gets metastable again
with two decay channels
\[
  (2Z,e,e) \rar (2Z,e) + e\ ,
\]
\[
  (2Z,e,e) \rar (Z,e) + (Z,e)\ .
\]
About the last domain we are not certain due to insufficient accuracy of our calculations.

A question to rise is a behavior of the total energy near the second critical point at $Z=0$ which is the singular point of the Schr\"odinger equation. Based on the fit of data from the domain $Z \in [0.1, 0.6]$ (five points, see Table IV) we find that the Puiseux expansion becomes the Taylor expansion
\[
   E_T\ =\ -0.6533 Z^2 - 4.1162 Z^3 + 2.5076 Z^4 + \ldots \ .
\]
Such a behavior does not provide an indication to singular nature of the point $Z=0$. However, the total energy can not be analytically continued to $\mbox{Re} Z <0$.

\subsection{$(3Z,e,e)$}

The system $(3Z,e,e)$ consists of three charged centers $Z$ and two electrons. For $Z=1$ it is celebrated H$_3^+$ molecular ion. It is obvious that for large $Z$ the system is unbound as well as for negative $Z$. Thus, there are two singular points: $Z=0$ where the potential "changes" sign and $Z_{cr}>2$ which is a critical point separating the domain of the existence from the domain of non-existence of the solutions in the Hilbert space. It seems natural that the ground state when exists is the spin-singlet state. Calculations (see below) show that $Z_{cr}=1.441$ at $R_{eq}=1.98$ a.u. The optimal geometrical configuration at equilibrium is the equilateral triangle. It was checked that this configuration is always stable with respect to small deviations. Thus, the system $(3Z,e,e)$ exists for charge $0<Z<Z_{cr}$.

\begin{figure}[h]
\includegraphics[angle=00, width=0.45\textwidth]{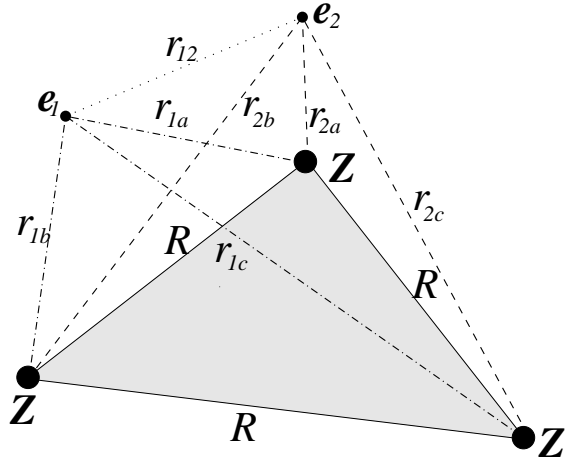}
\caption{Geometrical setting for $(3Z,e,e)$ system}
\end{figure}

{\it Trial Functions.} The variational method was used to obtain all numerical results. In general, the basic trial function $\psi^{(j)}$ has the form of a symmetrized product of six $1s$-Coulomb orbitals (Slater functions) and correlation function in exponential form (see \cite{TG:2007} for a discussion),
\begin{eqnarray}
\label{trial-3Z2e}
  \psi_g&=&e^{-\al_{1} r_{1a}-\al_{2} r_{1b}-\al_{3} r_{1c}-\al_{4} r_{2a}-\al_{5} r_{2b}-\al_{6} r_{2c}+\gamma r_{12}}\non\\
  &+&e^{-\al_{1} r_{1c}-\al_{2} r_{1a}-\al_{3} r_{1b}-\al_{4} r_{2c}-\al_{5} r_{2a}-\al_{6} r_{2b}+\gamma r_{12}}\non\\
  &+&e^{-\al_{1} r_{1b}-\al_{2} r_{1c}-\al_{3} r_{1a}-\al_{4} r_{2b}-\al_{5} r_{2c}-\al_{6} r_{2a}+\gamma r_{12}}\non\\
  &+&e^{-\al_{1} r_{1a}-\al_{2} r_{1c}-\al_{3} r_{1b}-\al_{4} r_{2a}-\al_{5} r_{2c}-\al_{6} r_{2b}+\gamma r_{12}}\non\\
  &+&e^{-\al_{1} r_{1c}-\al_{2} r_{1b}-\al_{3} r_{1a}-\al_{4} r_{2c}-\al_{5} r_{2b}-\al_{6} r_{2a}+\gamma r_{12}}\non\\
  &+&e^{-\al_{1} r_{1b}-\al_{2} r_{1a}-\al_{3} r_{1c}-\al_{4} r_{2b}-\al_{5} r_{2a}-\al_{6} r_{2c}+\gamma r_{12}}\non\\
  &+&e^{-\al_{1} r_{2a}-\al_{2} r_{2b}-\al_{3} r_{2c}-\al_{4} r_{1a}-\al_{5} r_{1b}-\al_{6} r_{1c}+\gamma r_{12}}\non\\
  &+&e^{-\al_{1} r_{2c}-\al_{2} r_{2a}-\al_{3} r_{2b}-\al_{4} r_{1c}-\al_{5} r_{1a}-\al_{6} r_{1b}+\gamma r_{12}}\non\\
  &+&e^{-\al_{1} r_{2b}-\al_{2} r_{2c}-\al_{3} r_{2a}-\al_{4} r_{1b}-\al_{5} r_{1c}-\al_{6} r_{1a}+\gamma r_{12}}\non\\
  &+&e^{-\al_{1} r_{2a}-\al_{2} r_{2c}-\al_{3} r_{2b}-\al_{4} r_{1a}-\al_{5} r_{1c}-\al_{6} r_{1b}+\gamma r_{12}}\non\\
  &+&e^{-\al_{1} r_{2c}-\al_{2} r_{2b}-\al_{3} r_{2a}-\al_{4} r_{1c}-\al_{5} r_{1b}-\al_{6} r_{1a}+\gamma r_{12}}\non\\
  &+&e^{-\al_{1} r_{2b}-\al_{2} r_{2a}-\al_{3} r_{2c}-\al_{4} r_{1b}-\al_{5} r_{1a}-\al_{6} r_{1c}+\gamma r_{12}}
\end{eqnarray}
It is worth noting that a linear superposition of three functions of a type
(\ref{trial-3Z2e}) leads to the most accurate energy for the H$_3^+$-molecule for lowest spin-triplet state in linear configuration ${}^3\Si_u$ among a few parametric trial functions giving a relative accuracy $\sim 10^{-3}$ \cite{TGL:2007}.

With such a function (\ref{trial-3Z2e}) the expected relative accuracy in total energy is $\approx 10^{-3}$. The total energy $E(Z,R=R_{eq})$ of $(3Z,e,e)$ is presented at Fig. \ref{E:nz+ee} and the equilibrium distance is at Fig. \ref{R:nz+ee}. The optimal geometrical configuration at equilibrium is always the equilateral triangle. It was checked that this configuration is always stable with respect to small deviations. Both curves are smooth without any indication to a special feature at the physical charge $Z=1$. At charge $Z=1.1767$ the energy curves for $(2Z,e,e)$ and $(3Z,e,e)$ intersect. This crossing separates the domain of stability from metastability of the system $(3Z,e,e)$: at $Z>1.1767$ $(3Z,e,e)$ can decay to $(2Z,e,e)+Z$.  The equilibrium distance $R_{eq}$ is a smooth curve which has a minimum $R_{eq}=1.643$ a.u. at $Z=0.8981$, it grows to infinity with a decrease of $Z$. Two equilibrium distances curves for $(2Z,e,e)$ and $(3Z,e,e)$ intersect twice for $Z=0.6851$ with $R_{eq}=1.6917$ a.u. and for $Z=0.3460$ with $R_{eq}=2.4297$ a.u.

Behavior of the energy as function of the charge $Z$ close to critical
charge $Z < Z_{cr}$ is given by the Puiseux expansion (see (\ref{Pui})):
\begin{eqnarray}
\label{z3+enfit}
  E(Z) & = & -3.6798+1.7613(Z_{cr}-Z)-0.5009(Z_{cr}-Z)^{3/2}\\\non
&+&1.5164(Z_{cr}-Z)^2+0.6143(Z_{cr}-Z)^{5/2} + \ldots \ ,
\end{eqnarray}
where the critical point is
\begin{equation}
\label{Z3+cr}
    Z_{cr}=1.441 \ .
\end{equation}
The fit (\ref{z3+enfit}) is based on data from the domain $Z\in[1.20,1.42]$ (7 points, see Table V). This behavior indicates that critical point seems to be a square-root branch point.

\begin{table}[htb]
\centering
\begin{tabular}{|c|c|c|}
\hline
 $Z$ &   $E_T$ & Fit\\\hline
0.30 & -0.3051 & -0.3058 \\\hline
0.40 & -0.5820 & -0.5809 \\\hline
0.50 & -0.9216 & -0.9224 \\\hline
0.60 & -1.2870 & -1.2868 \\\hline
\hline
1.20  & -3.2097 & -3.2095 \\\hline
1.25  & -3.3197 & -3.3206 \\\hline
1.30  & -3.4251 & -3.4237 \\\hline
1.35  & -3.5185 & -3.5196 \\\hline
1.40  & -3.6100 & -3.6094 \\\hline
1.41  & -3.6269 & -3.6268 \\\hline
1.42  & -3.6436 & -3.6440 \\\hline
\end{tabular}
\caption{Total energy $E_T$ of $(3Z,e,e)$ in Ry at equilibrium vs $Z$ obtained using (\ref{trial-3Z2e}) compared to the result of the fit (\ref{z3+enfit}).
}
\label{fit_3Z+2e}
\end{table}

There are three points of crossing for the energy curve $(3Z,e,e)$ displayed at Fig. \ref{E:nz+ee}.  The first one is $Z_{cross}^{(1)}=1.3566$ with $R_{eq}=1.881$ a.u. for the crossing of $(2Z,e)$ at $R_{eq}=2.406$ a.u. and the $(Z,e)$. The second one is  $Z_{cross}^{(2)}=1.3137$ with $R_{eq}=1.837$ a.u. for the crossing with two atomic $(Z,e)$ systems. The third one is  $Z_{cross}^{(3)}=1.1767$ with $R_{eq}=1.716$ a.u. for the crossing with the $(2Z,e,e)$ at $R_{eq}= 1.329$ a.u.

For charges $Z \in (1.3566,1.4407)$ the system is metastable with three decay channels
\begin{eqnarray}
(3Z,e,e) & \to & (2Z,e) + (Z,e)  \non\\
(3Z,e,e) & \to & 2(Z,e) + Z \non \\
(3Z,e,e) & \to & (2Z,e,e) + Z \non
\end{eqnarray}
For charges $Z \in (1.3137,1.3566)$ the system is metastable with two decay channels
\begin{eqnarray}
(3Z,e,e) & \to & 2(Z,e)+Z\non \\
(3Z,e,e) & \to & (2Z,e,e)+Z\non
\end{eqnarray}
For charges $Z\in (1.1767,1.3137)$ the system is metastable with one decay channel
\begin{eqnarray}
(3Z,e,e) & \to & (2Z,e,e) + Z \non
\end{eqnarray}
Eventually, for charges $Z\in (0.2989,1.1767)$ the system becomes stable.
For charges $Z < 0.2989$ the system can be either in the form $(3Z,e)+e$ or
$(2Z,e)+(Z,e)$. The accuracy of our calculations do not allow us to make a
definite statement.

Another question to rise is a behavior of the total energy near the second critical point at $Z_{cr}=0$ which is the singular point of the Schr\"odinger equation. Based on the fit of data from the domain $Z \in [0.1, 0.5]$ (five points, see Table V) we find that the Puiseux expansion becomes the Taylor expansion
\[
   E_T\ =\ -0.7198 Z^2 - 11.9676 Z^3 + 14.0751 Z^4 + \ldots \ .
\]
Such a behavior does not provide an indication to singular nature of the point $Z=0$. However, the total energy can not be analytically continued to $\mbox{Re} Z <0$.

\section*{Conclusions}

In this paper we calculated for the first time the critical charges of five simple 1-2 electron molecular systems: $(2Z,e), (3Z,e), (4Z,e), (2Z,e,e), (3Z,e,e)$ under the assumption that the $Z$-charges are static and found their equilibrium configurations. It was also found that for all those systems the total energy and equilibrium distance vs $Z$ are smooth curves without any indication to charge quantization. For all studied systems the optimal geometric configuration is the most symmetric being the equilateral triangle for  $(3Z,e), (3Z,e,e)$ and tetrahedron for $(4Z,e)$. It was checked that this configuration is always stable with respect to small deviations. It seems natural to assume that for $(4Z,e,e)$ the tetrahedron as the optimal geometrical configuration would occur. It would be interesting to study the optimal geometrical configuration for five (or more) $Z$-center cases, in particular, for $(5Z,e)$.

It is evident the existence of the critical charge for any one electron system $(nZ,e)$, since the potential has a form $V= -A Z + B Z^2$ with $A,B > 0$ and always becomes negative if the charge $Z$ is small enough. The critical charge behaves like $Z_{cr,n} \propto n^{\al}$ with some $\al < 0$ at large $n$. However, the question about stability of $(nZ,e)$ at $Z < Z_{cr,n}$ remains unclear to the present authors. Probably, a similar conclusion can be made for two electron systems.

\begin{acknowledgments}
  The authors are grateful to J. C. L\'opez Vieyra for helpful
  discussions, important assistance with computer calculations and for their
  interest in the present work. This work was supported in part by the
  university program FENOMEC and by the PAPIIT grant {\bf IN115709} and CONACyT grant {\bf 58962-F} (Mexico).
  H.M.C. thanks to PhD study support program through CONACyT grant {\bf 58962-F}.
\end{acknowledgments}

\end{document}